\documentclass[preprint,showpacs,preprintnumbers,amsmath,amssymb,showkeys]{revtex4}
 
\usepackage{graphicx}
\usepackage{dcolumn}
\usepackage{bm}
 
\begin{document} 
 
\noindent

\preprint{}

\title{Is nonlocality responsible for the violation of Bell's inequalities?}

\author{Agung Budiyono}
\email{agungbymlati@gmail.com} 

\affiliation{Jalan Emas 772 Growong Lor RT 04 RW 02 Juwana, Pati, 59185 Jawa Tengah, Indonesia}
 
\date{\today}
   
\begin{abstract} 
Bell's theorem has been widely argued to show that some of the predictions of quantum mechanics which are obtained by applying the {\it Born's rule} to a class of {\it entangled states}, are {\it not} compatible with {\it any} local-causal statistical model, via the violation of Bell's inequalities. On the other hand, in the previous work, we have shown that quantum dynamics and kinematics are {\it emergent} from a statistical model that is singled out {\it uniquely} by the principle of Locality. Here we shall show that the local-causal model supports entangled states and give the statistical origin of their generation. We then study the Stern-Gerlach experiment to show that the Born's rule can also be derived as a mathematical theorem in the local-causal model. These results lead us to argue that nonlocality is {\it not} responsible for the quantum mechanical and most importantly experimental violation of Bell's inequalities. The source(s) of violation has to be sought somewhere else.   
\end{abstract}   
   
\pacs{03.65.Ta, 03.65.Ud, 05.20.Gg}
\keywords{Statistical model of quantization; Bell's theorem; Entanglement; Born's rule; (Non)locality; (In)separability}
\maketitle  
 
\section{Motivation\label{motivation}}  
 
Attempts to provide a physically transparent, realist, and causal description underlying the abstract formalism and the inherent random nature of the predictions of quantum mechanics is as old as quantum mechanics itself. In this research program, one typically supplements the wave function with a set of hidden variables, traditionally denoted by $\lambda$, which together with the wave function determine completely the state of the system of interest, so that the Born's rule for the probability to get an outcome $\omega$ given the initial wave function of the system $\psi_0$ in an ensemble of identical measurements with setting parameters of the apparatus $b$, $P(\omega|b,\psi_0)$, admits causal description. The subscript `$0$' emphasizes that $\psi_0$ is the wave function of the system at the preparation. Within a hidden variable model, one therefore expects to decompose the probability to get various outcomes as a statistical mixture 
\begin{equation}
P(\omega|b,\psi_0)=\int d\lambda P(\omega|\lambda,b,\psi_0)P(\lambda|b,\psi_0), 
\label{pre Born's rule hvm SMQ} 
\end{equation}
such that $P(\lambda|b,\psi_0)$, the distribution of the hidden variables given the setting parameter and the wave function, and $P(\omega|\lambda,b,\psi_0)$, the probability to get various outcomes given the hidden variables for the same setting parameter and the wave function, are derivable as mathematical theorems with transparent physical meaning, rather than heuristically postulated as the Born's rule in standard quantum mechanics. The model is said to be deterministic if the outcome is determined by the hidden variables so that $P(\omega|\lambda,b,\psi_0)$ is either $1$ or $0$.     
  
Bell's theorem says that any statistical model of the above type is incompatible with the statistical predictions of quantum mechanics when the wave function of the system $\psi_0$ belongs to a class of entangled (inseparable) wave functions, if the model is local satisfying the separability hypothesis \cite{Bell paper,CHSH inequality,CH stochastic model,Bell's theorem general ES}. The latter means that in a spacelike separated pair of joint measurements as in Bell-type experiments, when the system is completely characterized, the probability to get a pair of outcomes $\omega_1$ and $\omega_2$ at the two wings of the experiment, when the apparatus setting are respectively $b_1$ and $b_2$, should be separable 
\begin{equation} 
P(\omega_1,\omega_2|\lambda,b_1,b_2,\psi_0)=P(\omega_1|\lambda,b_1,\psi_0)P(\omega_2|\lambda,b_2,\psi_0). 
\label{SH} 
\end{equation} 
Namely, the probability to obtain $\omega_{1(2)}$ at the first(second) wing must be independent of the apparatus setting $b_{2(1)}$ (parameter independence), and of the outcome $\omega_{2(1)}$ (outcome independence) of the measurement conducted at the second(first) wing \cite{Jarret PI and OI,Shimony PI and OI}, and any statistical correlation of the pair of outcomes must be describable in term of the past local common causes characterized completely by $\lambda$ and $\psi_0$. Note that while the above hypothesis is originally inspired by the locality of the theory of relativity which presumes a finite maximum velocity of interaction given by the velocity of light in vacuum, the no-signaling of the latter only requires the model to satisfy the parameter independence. In this sense, Eq. (\ref{SH}) is often referred to as Bell's locality assumption. In a deterministic model, $P(\omega_i|\lambda,b_i,\psi_0)$, $i=1,2$, are either $1$ or $0$ so that $\omega_i=\omega_i(\lambda,b_i)$, that is the outcome of measurement at one wing cannot depend on the setting parameter of the other wing, an assumption that is regarded by Bell as ``vital'' \cite{Bell paper}.   
  
There are other assumptions that are considered `crucial' for the derivation of Bell's theorem by some authors, but attract(ed) less attention of most workers in the field. First is the assumption of measurement independence which says that changing the setting parameter of the apparatus does not affect the distribution of the hidden variables that determine the measurement outcomes \cite{Bell free will,Shimony free will,Espagnat free will,Shimony model,Brans full causal,Kofler free will relaxation,Hall free will relaxation,Di Lorenzo MI,Vervoort}. Within the above form of statistical model, it means $P(\lambda|b,\psi_0)=P(\lambda|\psi_0)$ which is equivalent to, via Bayes theorem, $P(b|\lambda)=P(b)$. It is therefore often believed that it gives a mathematical representation of the freedom of the experimenter to choose the measurement setting at will independent of the hidden variables, thus is also referred to as freedom-to-choose or freewill hypothesis. Other authors have argued that Bell has tacitly made an assumption that there exists a single global/common Kolmogorov probability space (measure) for all pairs of {\it incompatible} measurement {\it contexts} corresponding to different settings which cannot be done simultaneously, which in general, as shown by Vorob'ev \cite{Vorobev theorem}, is {\it not} correct \cite{Accardi contextual loophole,Pitowsky contextual loophole,de la Pena contextual loophole,Khrennikov contextual loophole,Volovich contextual loophole,Hess-Philips,Hess contextual loophole,Nieuwenhuizen loopholes}.          
   
On the other hand, we have shown in the previous works that the abstract laws that govern the dynamics and kinematics of quantum mechanics can be derived from a statistical model based on a chaotic (effectively random) fluctuation of infinitesimal stationary action which is {\it not only} local satisfying the principle of Relativistic Locality, but is singled out {\it uniquely} by the latter \cite{AgungSMQ6,AgungSMQ8}. See also Refs. \cite{AgungSMQ4,AgungSMQ7} for related works. In the present paper we shall first argue that the model naturally supports entangled states: since the model is based on a random fluctuation of infinitesimal stationary action and action is evaluated in {\it configuration space} instead of in ordinary space, then for interacting compound system, the fluctuation of the whole can {\it not} be separated into the fluctuation of each subsystem. This {\it statistical inseparability} will be argued to generate entanglement during the interaction.   
 
We shall then apply the model to Stern-Gerlach experiment for the measurement of angular momentum. Identifying the configuration of the whole system+apparatus as the hidden variables which determine the measurement outcomes in the statistical model, we shall show that Eq. (\ref{pre Born's rule hvm SMQ}), with the whole wave function of the system+apparatus replacing $\psi_0$, leads immediately to the Born's rule. Born's rule is thus also shown as a mathematical theorem of the local-causal statistical model rather than as an independent postulate as in standard quantum mechanics. 

These results, that entanglement and Born's rule are emergent in the local-causal statistical model, allow us to argue that nonlocality is {\it not} responsible for the violation of Bell's inequality predicted by quantum mechanics and confirmed by numerous experiments \cite{Brunner review}. The source of violation must lie hiding somewhere else.  
 
\section{A local statistical model of microscopic randomness}

\subsection{A statistical model based on a random fluctuation of infinitesimal stationary action}

Let us first summarize the essential points of the statistical model of quantum fluctuation reported in Refs. \cite{AgungSMQ7,AgungSMQ4,AgungSMQ6,AgungSMQ8}. Let us assume that the Lagrangian depends on a parameter $\xi(t)$: $L=L(q,\dot{q};\xi)$, fluctuating sufficiently chaotic in a microscopic time scale, whose physical origin is not our present concern. Here $q$ is the configuration of the system, $\dot{q}\doteq dq/dt$, where $t$ is time parameterizing the evolution of the system. Let us assume that the time scale for the fluctuation of $\xi$ is $dt$. Let us then consider two infinitesimally close spacetime points $(q;t)$ and $(q+dq;t+dt)$ {\it in configuration space} such that $\xi$ is constant. Let us assume that fixing $\xi$, the principle of stationary action is valid to select a segment of path, denoted by $\mathcal{J}(\xi)$,  that connects the two points. One must then solve a variational problem with fixed end points: $\delta(Ldt)=0$. This variational problem leads to the existence of a function $A(q;t,\xi)$, the Hamilton's principal function, whose differential along the path is given by, for a fixed $\xi$, 
\begin{equation} 
dA=Ldt=p\cdot dq-Hdt, 
\label{infinitesimal stationary action} 
\end{equation} 
where $p(\dot{q})=\partial L/\partial{\dot{q}}$ is the momentum, and $H(q,p;\xi)\doteq p\cdot\dot{q}(p)-L(q,\dot{q}(p);\xi)$ is the Hamiltonian which is also parameterized by $\xi(t)$. Hence, $dA(\xi)$ is just the `infinitesimal stationary action' along the corresponding short path during the infinitesimal time interval $dt$ in which $\xi$ is fixed. Let us here note that we have implicitly assumed that $\xi$ is a global parameter, having a uniform value across the universe. This is clear for example when considering a system of two (or more) particles arbitrarily spatially separated from each other. It is therefore natural to assume that $\xi(t)$ is cosmological in origin the detail of which will be discussed somewhere else. 

Varying $\xi$, the principle of stationary action will therefore pick up various different paths $\mathcal{J}(\xi)$ {\it in configuration space}, all connecting the same two infinitesimally close spacetime points, each with different values of infinitesimal stationary action $dA(\xi)$. Due to the chaotic fluctuation of $\xi(t)$, $dA(\xi)$ is thus fluctuating effectively randomly. A single event description is therefore practically impossible and a statistical approach is inevitable. Hence, we have an effective stochastic processes in which the system starting with a configuration $q$ at time $t$ may take various different paths randomly to end up with a configuration $q+dq$ at time $t+dt$. The stochastic processes is therefore completely described by a `transition probability' for the system starting with a configuration $q$ at time $t$ to move to its infinitesimally close neighbor $q+dq$ at time $t+dt$ via a path $\mathcal{J}(\xi)$, denoted below by $P((q+dq;t+dt)|\{\mathcal{J}(\xi),(q;t)\})$. 
 
Let us further assume that the transition probability above is determined by the fluctuation of the infinitesimal stationary action according to the following exponential law \cite{AgungSMQ4,AgungSMQ6,AgungSMQ7,AgungSMQ8}:
\begin{eqnarray} 
P((q+dq;t+dt)|\{\mathcal{J}(\xi),(q;t)\})\hspace{25mm}\nonumber\\
\propto N \exp\Big({-\frac{2}{\gamma(\xi)}\Big[dS(q;t,\xi)-dA(q;t,\xi)\Big]}\Big)\nonumber\\
\doteq P(dS|dA), 
\label{exponential distribution of DISA}
\end{eqnarray}  
where $S(q;t,\xi)$ is a chaotic (effectively random) quantity whose differential along the path $\mathcal{J}(\xi)$ is given by 
\begin{equation}
dS(q;t,\xi)=\frac{dA(q;t,\xi)+dA(q;t,-\xi)}{2}=dS(q;t,-\xi),  
\label{infinitesimal phase}
\end{equation} 
$N$ is a factor independent of $dS-dA$ the form of which to be specified later and $\gamma(\xi)$ is a non-vanishing function of $\xi$ with action dimensional, thus is in general randomly fluctuating along with time. The justification of the above exponential law will be given later. From Eq. (\ref{infinitesimal phase}), one has $dS(q;t,\xi)-dA(q;t,\xi)=(dA(q;t,-\xi)-dA(q;t,\xi))/2$. Hence, to guarantee the negative definiteness of the exponent in Eq. (\ref{exponential distribution of DISA}) for normalizability, $\gamma$ must flip its sign as $\xi$ changes its sign. This fact allows us to assume that both $\gamma$ and $\xi$ always have the {\it same} sign. The time scale for the fluctuation of the sign of $\gamma$ must therefore be the same as that of $\xi$ given by $dt$. Since $dA$ is just the infinitesimal stationary action, we shall refer to Eq. (\ref{exponential distribution of DISA}) as the `distribution of the deviation from infinitesimal stationary action'. 

Let us denote the magnitude of the envelope of $\xi$ as $\|\xi\|$. It fluctuates in a time scale $\tau_{\xi}$, assumed to be much larger than the time scale of the fluctuation of the sign of $\xi$: $\tau_{\xi}\gg dt$. It is then clear that for the distribution of Eq. (\ref{exponential distribution of DISA}) to make sense mathematically, the time scale of the fluctuation of $|\gamma|$, denoted by $\tau_{\gamma}$, must be much larger than the time scale of the fluctuation of $\|\xi\|$. One thus has 
\begin{equation}
\tau_{\gamma}\gg\tau_{\xi}\gg dt. 
\label{time scales}
\end{equation}
Hence, in a time interval of length $\tau_{\xi}$, the magnitude of the envelope of $\xi$ is effectively constant while the sign of $\xi$ may fluctuate randomly together with the sign of $\gamma$ in a time scale $dt$. Moreover, in a time interval of length $\tau_{\gamma}$, $|\gamma|$ is effectively constant and $\|\xi\|$ fluctuates randomly so that the distribution of $|dS-dA|$ is given by the exponential law of Eq. (\ref{exponential distribution of DISA}) characterized by $|\gamma|$. Let us further assume that the probability density of the occurrence of $\xi$ at any time satisfies the following unbiased condition:  
\begin{equation}
P(\xi)=P(-\xi),  
\label{God's unbiased}
\end{equation}  
namely both of the signs of $\xi$ are equally likely to occur. This assumption is equivalent to the assumption that the two signs of $dS-dA$ occur equally probably. The probability density of $\gamma$ is therefore also unbiased $P(\gamma)=P(-\gamma)$. 

From the exponential distribution of Eq. (\ref{exponential distribution of DISA}), fixing $|\gamma|$ which is valid for a time interval less than $\tau_{\gamma}$, the average deviation from infinitesimal stationary action is given by 
\begin{equation}
\overline{dS-dA}=|\gamma|/2 . 
\label{average cosmological fluctuation}
\end{equation}
It is then natural to assume that the classical limit of macroscopic regime corresponds to the case when the average deviation from infinitesimal stationary action is much smaller than the infinitesimal stationary action itself: $|dA|/|\gamma|\gg 1$, or formally when $|\gamma|\rightarrow 0$. In this limiting case, from Eq. (\ref{exponential distribution of DISA}), one has $P(dS|dA)\rightarrow\delta(dS-dA)$, or $dS\rightarrow dA$, so that $S$ satisfies the Hamilton-Jacobi equation due to Eq. (\ref{infinitesimal stationary action}). This suggests that $|\gamma|$ must take microscopic values. 

Now let $\Omega(q,\xi;t)$ denotes the joint-probability density that the configuration of the system is $q$ and a random value of $\xi$ is realized at time $t$. Then according to the conventional (classical) probability theory, fixing $\xi$, the joint-probability density that the system initially at $(q;t)$ traces the segment of trajectory $\mathcal{J}(\xi)$ and end up at $(q+dq;t+dt)$, denoted below as $\Omega\big(\{(q+dq,\xi;t+dt),(q,\xi;t)\}\big|\mathcal{J}(\xi)\big)$, is equal to the probability density that the configuration of the system is $q$ at time $t$, $\Omega(q,\xi;t)$, multiplied by the transition probability between the two infinitesimally close points via the segment of trajectory $\mathcal{J}(\xi)$ which is given by Eq. (\ref{exponential distribution of DISA}). One thus has  
\begin{eqnarray}
\Omega\Big(\{(q+dq,\xi;t+dt),(q,\xi;t)\}\big|\mathcal{J}(\xi)\Big)\hspace{10mm}\nonumber\\
= P((q+dq;t+dt)|\{\mathcal{J}(\xi),(q;t)\})\times\Omega(q,\xi;t)\nonumber\\
\propto Ne^{-\frac{2}{\gamma}(dS(\xi)-dA(\xi))}\times\Omega(q,\xi;t).  
\label{probability density} 
\end{eqnarray}    

To further elaborate the above equation, one has to know the functional form of $N$. Let us assume that $N$ takes the following general form:
\begin{equation}
N\propto\exp(-\theta(S)dt), 
\label{exponential classical}
\end{equation}
where $\theta$ is a scalar function of $S$. Further let us consider the case when $|(dS-dA)/\gamma|\ll 1$. Inserting Eq. (\ref{exponential classical}) into Eq. (\ref{probability density}) and expanding the exponential on the right hand side up to the first order one gets
\begin{eqnarray}
d\Omega=-\Big[\frac{2}{\gamma}(d S-dA)+\theta(S)d t\Big]\Omega,    
\label{fundamental equation 0}
\end{eqnarray} 
where $d\Omega(q,\xi;t)\doteq\Omega\big(\{(q+dq,\xi;t+dt),(q,\xi;t)\}\big|\mathcal{J}(\xi)\big)-\Omega(q,\xi;t)$ is the change of the probability density $\Omega$ due to the transport along the segment of trajectory $\mathcal{J}(\xi)$. 

To guarantee a smooth correspondence with classical mechanics, the above equation must describe the dynamics of the classical ensemble of trajectories when $dS=dA$. Putting $dS=dA$ in Eq. (\ref{fundamental equation 0}), dividing both sides by $dt$ and taking the limit $dt\rightarrow 0$, one obtains $\dot{\Omega}+\theta(A)\Omega=0$. This last equation must therefore be identified as the continuity equation for the ensemble of classical trajectories. To do this, it is sufficient to choose $\theta(S)$ to be determined uniquely by the classical Hamiltonian as \cite{AgungSMQ6,AgungSMQ4,AgungSMQ7,AgungSMQ8}  
\begin{equation}
\theta(S)=\partial_q\cdot\Big(\frac{\partial H}{\partial p}\Big|_{p=\partial_qS}\Big), 
\label{QC correspondence}
\end{equation}
so that for $dS=dA$, it is given by the divergence of the corresponding classical velocity field.  
 
Next, recalling that $\xi$ is fixed during the infinitesimal time interval $dt$, one can expand the differentials $d\Omega$ and $dS$ in Eq. (\ref{fundamental equation 0}) as $dF=\partial_tF dt+\partial_qF\cdot dq$. Using Eq. (\ref{infinitesimal stationary action}), one finally obtains the following pair of coupled differential equations:  
\begin{eqnarray}
p(\dot{q})=\partial_qS+\frac{\gamma}{2}\frac{\partial_q\Omega}{\Omega},\hspace{15mm}\nonumber\\
-H(q,p(\dot{q});\xi)=\partial_tS+\frac{\gamma}{2}\frac{\partial_t\Omega}{\Omega}+\frac{\gamma}{2}\theta(S). 
\label{fundamental equation rederived}
\end{eqnarray}
 
It is evident that, as expected, in the formal limit $|\gamma|\rightarrow 0$, Eq. (\ref{fundamental equation rederived}) reduces to the classical Hamilton-Jacobi equation. In this sense, Eq. (\ref{fundamental equation rederived}) can be regarded as a natural generalization of the Hamilton-Jacobi equation due to the random deviation from infinitesimal stationary action following the exponential law of Eq. (\ref{exponential distribution of DISA}). Unlike the Hamilton-Jacobi equation in which we have a single unknown (Hamilton's principal) function $A$, however, to calculate the velocity or momentum and energy, one now needs a pair of unknown functions $S$ and $\Omega$. The relations in Eq. (\ref{fundamental equation rederived}) must {\it not} be interpreted that the momentum and energy of the particles are determined causally by the gradient of the probability density $\Omega$ (or $\ln(\Omega)$), which is physically absurd, rather it is the other way around as described explicitly by Eq. (\ref{probability density}).   

Let us end this subsection by listing a couple of symmetry relations for $S$ and $\Omega$ in the model which are valid by construction. First, from Eq. (\ref{infinitesimal phase}) one obtains, for a fixed value of $\xi$, the following symmetry relations:  
\begin{eqnarray}
\partial_qS(q;t,\xi)=\partial_qS(q;t,-\xi),\nonumber\\
\partial_tS(q;t,\xi)=\partial_tS(q;t,-\xi). 
\label{quantum phase symmetry}
\end{eqnarray}
Moreover, to comply with Eq. (\ref{God's unbiased}), $\Omega(q,\xi;t)$ has to satisfy the following symmetry relation:
\begin{eqnarray} 
\Omega(q,\xi;t)=\Omega(q,-\xi;t). 
\label{God's fairness}
\end{eqnarray}  
Both Eqs. (\ref{quantum phase symmetry}) and (\ref{God's fairness}) will play important roles later.

\subsection{Principle of Locality and statistical inseparability\label{locality and inseparability}}
  
Let us proceed to show that the above statistical model is consistent with the locality hypothesis demanded by the theory of relativity that due to the finite maximum velocity of interaction given by the velocity of light in vacuum, two subsystems, one is outside the light cone of the other, cannot influence each other. To see this, it is sufficient to consider a compound system composed of two particles whose configuration is denoted by $q=(q_1,q_2)$, sufficiently separated from each other so that due to the relativistic locality, there is no physical-mechanical interaction between the two. The Lagrangian is thus decomposable as $L(q_1,q_2,\dot{q}_1,\dot{q}_2;\xi)=L_1(q_1,\dot{q}_1;\xi)+L_2(q_2,\dot{q}_2;\xi)$, so that the infinitesimal stationary action is also decomposable: $dA(q_1,q_2;\xi)=dA_1(q_1;\xi)+dA_2(q_2;\xi)$, and accordingly one has, by virtue of Eq. (\ref{infinitesimal phase}), $dS(q_1,q_2;\xi)=dS_1(q_1;\xi)+dS_2(q_2;\xi)$. On the other hand, since the Hamiltonian $H$ is decomposable as $H(q_1,q_2,p_1,p_2;\xi)=H_1(q_1,p_1;\xi)+H_2(q_2,p_2;\xi)$, $p_{i}$, $i=1,2$, is the momentum of the $i-$particle, then $\theta$ of Eq. (\ref{exponential classical}) is also decomposable: $\theta(q_1,q_2;\xi)=\theta_1(q_1;\xi)+\theta_2(q_2;\xi)$, so that $N$ is separable $N(q_1,q_2;\xi)=N_1(q_1;\xi)N_2(q_2;\xi)$.  
 
Inserting all these into Eqs. (\ref{exponential distribution of DISA}) and (\ref{exponential classical}), one can explicitly see that the distribution of deviation from infinitesimal stationary action for the two spacelike separated particles is {\it separable} as 
\begin{equation}
P(dS_1+dS_2|dA_1+dA_2)=P(dS_1|dA_1)P(dS_2|dA_2).   
\label{principle of Local Causality}
\end{equation}
In this case, $P(dS|dA)=P(dS_1+dS_2|dA_1+dA_2)$ can thus be regarded as the joint-probability distribution of the deviation from infinitesimal stationary action of the non-interacting two particles system, and moreover, it is {\it separable} into the probability distribution of the deviation with respect to each single particle. Each is therefore {\it independent} of the other as required by the principle of Relativistic Locality, that is what happens with one of the particles has {\it no} effect whatsoever on the dynamics and statistics of the other spacelike separated particle. One can also clearly see that unlike the separability hypothesis of Eq. (\ref{SH}) which refers to the statistics of outcomes of measurement, the separability of Eq. (\ref{principle of Local Causality}) is {\it objective} referring directly to the {\it real factual} situation (state) of the system, independent of measurement. 

Notice further that the statistical separability for spacelike separated subsystems described by Eq. (\ref{principle of Local Causality}) is {\it unique} to the exponential law. A Gaussian distribution of deviation from infinitesimal stationary action for example does not have such a property. Namely, for a Gaussian law, the spacelike separated pair of particles may still influence each other in contradiction with the principle of Locality of the theory of relativity. Noting this fact, we have argued in Refs. \cite{AgungSMQ6,AgungSMQ8} that the exponential law for the transition probability of Eq. (\ref{exponential distribution of DISA}) is the {\it unique} form of the probability distribution of the deviation from infinitesimal stationary action, up to the global parameter $\gamma$, that is {\it singled out} among the infinitude of possibilities by the principle of Locality mathematically represented by Eq. (\ref{principle of Local Causality}).  
 
Now let us proceed to consider the case when the two particles compound system are {\it interacting} with each other. In this case, the total Lagrangian $L$ is {\it not} decomposable so that the infinitesimal stationary action $dA$ and also $dS$, which are evaluated in configuration space instead of in ordinary space, are not decomposable either. Equation (\ref{principle of Local Causality}) does no more hold. $P(dS|dA)$ can {\it no} longer be regarded as a joint-probability density of the fluctuation of the deviation from infinitesimal stationary action with respect to the two particles. One cannot attribute to each particle a probability density of the fluctuation of the deviation from infinitesimal stationary action. The only way to identify the statistics of the fluctuation of infinitesimal stationary action of the system is through $P(dS|dA)$ which, for interacting compound systems, refers to the {\it whole} compound. 

In this sense, there is a {\it statistical inseparability} of the randomness in the whole system: the randomness of the whole compound system, which is generated by the fluctuation of infinitesimal stationary action, can{\it not} be regarded as arising from the randomness of each subsystem. Rather it is the other way around: the randomness of each subsystem is induced by the random fluctuation of the whole compound system or {\it the randomness of the whole gives a global constraint/context to the randomness of each subsystem}. We have thus a top-down information flow \cite{Ellis top-down}. Statistically, the interacting compound system must therefore be regarded as a single unanalyzable (indivisible) whole. Note however that one can still identify each particle by specifying its position and momentum.  

This situation for interacting compound system in the statistical model is fundamentally different from the conventional stochastic motion based on a pair of random forces, each acting on a single particle. In this later case, one can always define joint-probability density of the two random forces so that the randomness of the whole compound system can be said to originate from the combination of the randomness of each subsystem and their correlation via the usual bottom-up information flow. We shall show in the next sections that the above statistical inseparability in the statistical model due to a top-down information flow is responsible for the so-called quantum mechanical entanglement, and also for the peculiar properties of quantum measurement which are absent in the conventional classical mechanics. 

\section{Quantization: the Stern-Gerlach experiment}
 
Let us apply, in this section, the statistical model of microscopic random deviation from classical mechanics developed in the previous section to investigate the Stern-Gerlach experiment for the measurement of angular momentum. We shall first discuss in the next subsection the Stern-Gerlach experiment within classical mechanics, and apply the statistical model to `quantize' its dynamics in the following subsection. 

\subsection{Classical model of Stern-Gerlach experiment\label{classical SG}}

Let us assume that we have a beam of neutral atoms whose center of mass coordinate is denoted by $q_a=(x_a,y_a,z_a)$, each containing an electron with coordinate $q_e=(x_e,y_e,z_e)$, sent one by one to a Stern-Gerlach apparatus. The interaction between the atom and the magnetic field of the Stern-Gerlach apparatus is thus mainly due to the angular momentum of the electron $l_e=q_e\times p_e$. Further, let us assume that the magnetic field is non-vanishing only in one direction and take that direction as the $z-$axis: $B=(0,0,B_z)$. For the sake of simplicity, let us assume that $B_z$ is linear in $z$: $B_z=B'z_a$, where $B'$ is some constant \cite{some complication}. In this case, the classical interaction-Hamiltonian can be approximated as
\begin{equation} 
H_I=\frac{e}{2m_e c}B\cdot l_e\approx\mu z_a l_{z_e}=\mu z_a(x_ep_{y_e}-y_ep_{x_e}),
\label{classical Hamiltonian SG angular momentum C}
\end{equation}
where $\mu=eB'/2m_ec$ with $e$ is the charge of electron, $m_e$ is its mass, $c$ is the velocity of light, and $l_{z_e}=x_ep_{y_e}-y_ep_{x_e}$ is the $z-$component of the angular momentum of the electron. 

Further, let us assume that the free Hamiltonian of the electron is negligible as compared to that of the rest of the atom. Hence, the total Hamiltonian is approximately given by 
\begin{equation}
H\approx H_I+H_a. 
\end{equation}
Here $H_a$ is the free Hamiltonian of the atom which reads 
\begin{equation}
H_a=\frac{p_a^2}{2m_a},
\label{free atomic Hamiltonian}
\end{equation}
where $m_a$ is the mass of the atom. 

First, one has, during the measurement-interaction $\dot{l}_{z_e}=\{l_{z_e},H\}=0$. Hence, the measurement-interaction conserves the $z-$angular momentum of the electron {\it prior} to interaction. The other component of angular momentum are in general not conserved. On the other hand, one also has $\dot{p}_{z_a}=\{p_{z_a},H\}=\mu l_{z_e}$, which can be integrated twice, noting that $l_{z_e}$ is a constant of motion, to give  
\begin{equation}
\dot{z}_a(T)=\frac{\mu}{m_a}l_{z_e}T,\hspace{2mm}z_a(T)=\frac{1}{2}\frac{\mu}{m_a}l_{z_e}T^2,   
\label{velocity and position post-interaction}
\end{equation} 
where $T$ is the time-span of measurement-interaction and we have used $p_a=m_a\dot{q}_a$ and assumed that $\dot{z}_a(0)=z_a(0)=0$. 

At time $t\ge T$ the particle leaves the magnetic region of the Stern-Gerlach apparatus so that the atom now moves approximately freely dictated by the free Hamiltonian $H_a$ with the initial velocity and position given by Eq. (\ref{velocity and position post-interaction}). Let us define $t_M=t-T$. The $z-$coordinate of the atom at $t\ge T$ then evolves with time as 
\begin{equation} 
z_a(t_M)=\frac{\mu}{m_a}l_{z_e}Tt_M+z_a(T). 
\label{classical pointer} 
\end{equation}
In this way $z_a(t_M)$ is naturally regarded as the pointer of measurement (the reading of the experiment) from the value of which one can {\it operationally infer} the value of $l_{z_e}$ {\it prior} to measurement. This is in principle what is actually done in all experiments, either involving macroscopic or microscopic objects, where one reads the position of the needle of the meter or the position of the detector that `clicks', etc: all experiments ultimately reduce to the determination of position of pointer. 

It is then natural to keep the above operationally clear measurement mechanism as we proceed in the next subsection to subject the classical system to the chaotic (effectively random) fluctuation of infinitesimal stationary action according to the statistical model of the previous section. Let us emphasize again that in the classical mechanical model of the Stern-Gerlach experiment, it is the angular momentum of the electron {\it prior} to measurement-interaction which determines the final value of the pointer (the position of the atom). We shall show in the next section that in the statistical model, due to the random fluctuation of infinitesimal stationary action of the whole system, this is no longer the case: the setting of the Stern-Gerlach apparatus will play irreducible role.   

\subsection{Statistical model of Stern-Gerlach experiment: wave function, Schr\"odinger equation, Hermitian operators and Born's statistical interpretation}

Let us modify the above classical mechanical model of Stern-Gerlach experiment according to the statistical model developed in the previous section. First, one needs to put a chaotic parameter $\xi(t)$ into the classical Hamiltonian of Eqs. (\ref{classical Hamiltonian SG angular momentum C}) and (\ref{free atomic Hamiltonian}) so that the infinitesimal stationary action is randomly fluctuating according to the exponential law of Eq. (\ref{exponential distribution of DISA}). Let us therefore assume that all the masses of the electron and the atom depend on $\xi(t)$, denoted respectively by $m_e^{\xi}$ and $m_a^{\xi}$. The total Hamiltonian thus reads
\begin{equation}
H(q,p;\xi)=\mu_{\xi}z_al_{z_e}+\frac{p_a^2}{2m_a^{\xi}}.
\label{total Hamiltonian SG angular momentum}
\end{equation}
where $\mu_{\xi}=eB'/2m_e^{\xi}c$. Let us further assume that $m_{e(a)}^{\xi}$ are fluctuating around their classical values $m_{e(a)}$ as 
\begin{eqnarray}
m_{e(a)}^{\xi}=m_{e(a)}+f_{e(a)}(\xi),\hspace{20mm}\nonumber\\
\mbox{with}\hspace{2mm} f_{e(a)}(\xi)=-f_{e(a)}(-\xi),\hspace{2mm}\mbox{and}\hspace{2mm}|f_{e(a)}(\xi)|\sim o(|\xi|). 
\label{fluctuation of metric}
\end{eqnarray}

Let us then consider a time interval of length $\tau_{\xi}$ during which $\|\xi\|$ is effectively constant while the sign of $\xi$ fluctuates randomly with equal probability so that the pair of equations in (\ref{fundamental equation rederived}) applies. Note that from Eq. (\ref{time scales}), in this interval of time, $|\gamma|$ is also effectively constant. Using the kinematic part of the Hamilton equation, $\dot{q}=\partial H/\partial p$, inserting Eq. (\ref{total Hamiltonian SG angular momentum}), the upper equation of (\ref{fundamental equation rederived}) reads 
\begin{eqnarray}
\dot{x}_e=-g_{\xi}(z_a)y_e,\hspace{2mm}\dot{y}_e=g_{\xi}(z_a)x_e,\hspace{2mm}\dot{z}_e=0,\nonumber\\
\dot{q}_a=\frac{\partial_{q_a}S}{m_a^{\xi}}+\frac{\gamma}{2m_a^{\xi}}\frac{\partial_{q_a}\Omega}{\Omega}, \hspace{20mm}
\label{classical velocity field SG angular momentum}
\end{eqnarray}
where $g_{\xi}(z_a)\doteq \mu_{\xi} z_a$. Assuming the conservation of probability, one then obtains the following continuity equation:
\begin{eqnarray}
0=\partial_t\Omega+\partial_q\cdot(\dot{q}\Omega)\hspace{40mm}\nonumber\\
=\partial_t\Omega+g_{\xi}(z_a)(x_e\partial_{y_e}\Omega-y_e\partial_{x_e}\Omega)+\partial_{q_a}\cdot\Big(\frac{\partial_{q_a}S}{m_a^{\xi}}\Omega\Big)\nonumber\\
+\frac{\gamma}{2m_a^{\xi}}\partial_{q_a}^2\Omega. \hspace{20mm}
\label{continuity equation SG angular momentum}
\end{eqnarray}

On the other hand, from Eq. (\ref{total Hamiltonian SG angular momentum}), $\theta(S)$ of Eq. (\ref{QC correspondence}) is given by 
\begin{equation}
\theta(S)=\partial_{q_a}^2S/m_a^{\xi}. 
\end{equation} 
The lower equation of (\ref{fundamental equation rederived}) thus becomes
\begin{eqnarray}
-H(q,p(\dot{q}))=\partial_tS+\frac{\gamma}{2}\frac{\partial_t\Omega}{\Omega}+\frac{\gamma}{2m_a^{\xi}}\partial_{q_a}^2S.
\label{fundamental equation particle in potentials} 
\end{eqnarray}
Plugging the upper equation of (\ref{fundamental equation rederived}) into the left hand side of Eq. (\ref{fundamental equation particle in potentials}), and using Eq. (\ref{total Hamiltonian SG angular momentum}), one obtains, after arrangement
\begin{eqnarray}
\partial_tS+g_{\xi}(z_a)(x_e\partial_{y_e}S-y_e\partial_{x_e}S)+\frac{(\partial_{q_a}S)^2}{2m_a^{\xi}}\hspace{20mm}\nonumber\\
-\frac{\gamma^2}{2m_a^{\xi}}\frac{\partial_{q_a}^2R}{R}
+\frac{\gamma}{2\Omega}\Big(\partial_t\Omega+g_{\xi}(z_a)(x_e\partial_{y_e}\Omega-y_e\partial_{x_e}\Omega)\nonumber\\
+\partial_{q_a}\cdot\Big(\frac{\partial_{q_a}S}{m_a^{\xi}}\Omega\Big)+\frac{\gamma}{2m_a^{\xi}}\partial_{q_a}^2\Omega\Big)=0, 
\label{gagego}
\end{eqnarray} 
where $R\doteq\sqrt{\Omega}$ and we have used the identity:
\begin{equation}
\frac{1}{4}\frac{\partial_{q_i}\Omega}{\Omega}\frac{\partial_{q_j}\Omega}{\Omega}=\frac{1}{2}\frac{\partial_{q_i}\partial_{q_j}\Omega}{\Omega}-\frac{\partial_{q_i}\partial_{q_j}R}{R}. 
\label{fluctuation decomposition}
\end{equation} 
Taking into account Eq. (\ref{continuity equation SG angular momentum}), the last term in the bracket of Eq. (\ref{gagego}) vanishes to give 
\begin{equation}
\partial_tS+g_{\xi}(z_a)(x_e\partial_{y_e}S-y_e\partial_{x_e}S)+\frac{(\partial_{q_a}S)^2}{2m_a^{\xi}}-\frac{\gamma^2}{2m_a^{\xi}}\frac{\partial_{q_a}^2R}{R}=0.
\label{HJM equation SG angular momentum}
\end{equation}

We have thus a pair of Eqs. (\ref{continuity equation SG angular momentum}) and (\ref{HJM equation SG angular momentum}) which is parameterized by $\gamma$. This pair of equations is valid in a microscopic time interval of length $\tau_{\xi}$ during which the magnitude of the envelope of $\xi$, that is $\|\xi\|$, is constant while its sign changes randomly with equal probability \cite{AgungSMQ6,AgungSMQ7,AgungSMQ8}. Averaging Eq. (\ref{continuity equation SG angular momentum}) for the cases $\pm\xi$, recalling that the sign of $\gamma$ is the same as that of $\xi$, one has, by virtue of Eqs. (\ref{quantum phase symmetry}) and (\ref{God's fairness}), 
\begin{eqnarray}
\partial_t\Omega+\widetilde{g}_{\xi}(z_a)(x_e\partial_{y_e}\Omega-y_e\partial_{x_e}\Omega)+\widetilde{1/m_a^{\xi}}\partial_{q_a}\cdot\Big(\partial_{q_a}S\Omega\Big)+\Delta(1/m_a^{\xi})\frac{|\gamma|}{2}\partial_{q_a}^2\Omega=0,
\label{QCE for angular momentum measurement approx}
\end{eqnarray}  
where for any function $H$ of $\xi$, we have defined
\begin{eqnarray}
\widetilde{H}=\frac{1}{2}(H(\xi)+H(-\xi)),\hspace{2mm}\Delta H=\frac{1}{2}(H(\xi)-H(-\xi)). 
\label{linear approx}
\end{eqnarray}
Similarly, averaging Eq. (\ref{HJM equation SG angular momentum}) over the cases $\pm\xi$, thus is also over $\pm\gamma$, one gets 
\begin{equation}
\partial_tS+\widetilde{g}_{\xi}(z_a)(x_e\partial_{y_e}S-y_e\partial_{x_e}S)+\widetilde{(1/m_a^{\xi})}\Big(\frac{(\partial_{q_a}S)^2}{2}-\frac{\gamma^2}{2}\frac{\partial_{q_a}^2R}{R}\Big)=0.
\label{HJM equation SG angular momentum approx}
\end{equation}
We thus have a pair of Eqs. (\ref{QCE for angular momentum measurement approx}) and (\ref{HJM equation SG angular momentum approx}) which are now parameterized by $|\gamma|$ valid for a microscopic time interval of length $\tau_{\xi}$ characterized by a constant $\|\xi\|$.  

Next, from Eq. (\ref{fluctuation of metric}), assuming that the fluctuation of $\xi$ around its vanishing average is sufficiently narrow, one has, in the lowest order approximation 
\begin{eqnarray}
\widetilde{g}_{\xi}(z_a)\approx \mu z_a\doteq g(z_a), \hspace{2mm} \widetilde{1/m_a^{\xi}}\approx 1/m_a,\hspace{2mm}\Delta(1/m_a^{\xi})\approx 0.
\end{eqnarray}
Taking this into account, the pair of Eqs. (\ref{QCE for angular momentum measurement approx}) and (\ref{HJM equation SG angular momentum approx}) can thus be approximated as
\begin{eqnarray}
\partial_t\Omega+g(z_a)(x_e\partial_{y_e}\Omega-y_e\partial_{x_e}\Omega)+\partial_{q_a}\cdot\Big(\frac{\partial_{q_a}S}{m_a}\Omega\Big)=0,\nonumber\\
\partial_tS+g(z_a)(x_e\partial_{y_e}S-y_e\partial_{x_e}S)+\frac{(\partial_{q_a}S)^2}{2m_a}-\frac{\gamma^2}{2m_a}\frac{\partial_{q_a}^2R}{R}=0.
\label{pair HJM equation SG angular momentum approx}
\end{eqnarray}
Further, noting that $\gamma$ is non-vanishing, let us proceed to define the following complex-valued function:
\begin{equation}
\Psi\doteq \sqrt{\Omega}\exp\Big(i\frac{S}{|\gamma|}\Big). 
\label{general wave function}  
\end{equation} 
Using $\Psi$ and recalling that $|\gamma|$ is constant during the microscopic time interval of interest with length $\tau_\xi$, the pair of equations in (\ref{pair HJM equation SG angular momentum approx}) can then be recast into the following compact form: 
\begin{equation}
i|\gamma|\partial_t\Psi=g(z_a)\big(-ix_e|\gamma|\partial_{y_e}+iy_e|\gamma|\partial_{x_e}\big)\Psi-\frac{\gamma^2}{2m_a}\partial_{q_a}^2\Psi. 
\label{generalized Schroedinger equation angular momentum SG}
\end{equation} 

Let us then consider a specific case when $|\gamma|$ is given by the reduced Planck constant $\hbar$, namely $\gamma=\pm\hbar$ with equal probability for all the time, $\tau_{\gamma}=\infty$, so that the average of the deviation from infinitesimal stationary action distributed according to the exponential law of Eq. (\ref{exponential distribution of DISA}) is given by 
\begin{equation} 
\hbar/2. 
\label{P}
\end{equation}
Moreover, recalling that the fluctuation of $\xi$ around its vanishing average is sufficiently narrow, one can approximate $\Omega(q,\xi;t)$ and $S(q;t,\xi)$ by the corresponding zeroth order terms in their Taylor series, denoted respectively by $\rho_Q(q;t)$ and $S_Q(q;t)$. In this specific case, the lowest order approximation of Eq. (\ref{generalized Schroedinger equation angular momentum SG}) then gives the following Schr\"odinger equation:
\begin{eqnarray} 
i\hbar\partial_t\Psi_Q=(\hat{H}_I+\hat{H}_a)\Psi_Q,\hspace{0mm}\nonumber\\
\mbox{with}\hspace{2mm}\hat{H}_I=g(z_a)\hat{l}_{z_e},\hspace{2mm}\hat{H}_a=\frac{\hat{p}_a^2}{2m_a}, 
\label{Schroedinger equation SG angular momentum}
\end{eqnarray}
where $\hat{p}_i\doteq -i\hbar\partial_{q_i}$ is the quantum mechanical linear momentum operator corresponding to the $i-$degree of freedom, ${\hat l}_{z_e}\doteq x_e\hat{p}_{y_e}-y_e\hat{p}_{x_e}$ is the $z-$component quantum mechanical angular momentum operator pertaining to the electron, and the quantum mechanical wave function $\Psi_Q$ is defined as
\begin{equation}
\Psi_Q(q;t)\doteq\sqrt{\rho_Q(q;t)}e^{\frac{i}{\hbar}S_Q(q;t)}. 
\label{quantum wave function}
\end{equation} 
From Eq. (\ref{quantum wave function}), the Born's statistical interpretation of wave function is clearly valid by construction
\begin{equation}
\rho_Q(q;t)=|\Psi_Q(q;t)|^2.
\label{Born's statistical interpretation}  
\end{equation} 

Further, recall that Eq. (\ref{classical velocity field SG angular momentum}) is valid for a time interval of length $\tau_{\xi}$ in which $\|\xi\|$ is effectively constant while the sign of $\xi$ fluctuates randomly with equal probability. It is then natural to define an `effective' velocity as $\widetilde{\dot{q}}(|\xi|)=(\dot{q}(\xi)+\dot{q}(-\xi))/2$. Inserting Eq. (\ref{classical velocity field SG angular momentum}) one gets, noting that the sign of $\gamma$ is always the same as that of $\xi$,
\begin{eqnarray}
\widetilde{\dot{x}}_e=-g(z_a)y_e,\hspace{2mm}\widetilde{\dot{y}}_e=g(z_a)x_e,\hspace{2mm}\widetilde{\dot{z}}_e=0,\nonumber\\
\widetilde{\dot{q}}_a=\frac{\partial_{q_a}S_Q}{m_a}, \hspace{20mm}
\label{effective classical velocity field SG angular momentum}
\end{eqnarray}
where we have used Eqs. (\ref{quantum phase symmetry}) and (\ref{God's fairness}) and counted only the zeroth order terms.  

\subsection{Statistical inseparability as the origin of quantum entanglement\label{entanglement}}

As discussed at the end of Subsection \ref{locality and inseparability}, for a compound system with interacting subsystems, say a system of interacting two particles, the infinitesimal stationary action $dA$ is not decomposable so that $dS$ is {\it not} decomposable either. Since by construction $S$ is the phase of the wave function defined in Eqs. (\ref{general wave function}) or (\ref{quantum wave function}), then in this case, the phase must be in general {\it not} decomposable. Otherwise, if $S$ is decomposable, then its differential must also be decomposable so that the two particles must not be interacting with each other in contradiction with our initial assumption. Hence, for interacting two particles system, the wave function of the whole compound is in general {\it inseparable} into the wave function of each subsystem 
\begin{equation}
\Psi_Q(q_1,q_2)\neq\Psi_{Q_1}(q_1)\Psi_{Q_2}(q_2).  
\end{equation}
In other words, in this case, one can {\it not} attribute a wave function defined as in Eqs. (\ref{general wave function}) or (\ref{quantum wave function}) to each subsystem. 

We have thus an {\it entangled} wave function. Moreover, the inseparability of the wave function or entanglement is preserved even when the interaction is turned off. We shall give in the next Section an example of such a generation of entanglement and its preservation when the interaction is turned off while applying the statistical model to describe the Stern-Gerlach experiment. Let us note however that when the interaction is turned off, while the wave function is still inseparable, the statistics of infinitesimal change of its phase must satisfy the separability condition of Eq. (\ref{principle of Local Causality}) in accord with the principle of Locality. 

We have therefore argued that the statistical origin of the generation of the entanglement could be traced to the fact that since action is evaluated in configuration space instead of in ordinary space, then the fluctuation of the infinitesimal stationary action with respect to the whole compound system with interacting subsystems cannot be fundamentally separated into the fluctuation with respect to each subsystem. This further implies that the randomness of the whole compound induced by the chaotic fluctuation of infinitesimal stationary action cannot be regarded as to arise from the combination of the randomness of each subsystem and their correlation. Rather, as is already mentioned in Subsection \ref{locality and inseparability}, it is the other way around: the randomness of the whole gives a global constraint (context) to the randomness of each subsystem, a manifestation of a top-down causation \cite{Ellis top-down}. Each subsystem can{\it not} be truly random, but is correlated in a subtle and nontrivial way with the other.    

\section{Quantum measurement}
 
\subsection{A single measurement event}
 
Now let us discuss the detail process of a single measurement event in the Stern-Gerlach experiment. Let us first assume that the magnetic field is sufficiently strong so that the interaction is impulsive. In this case, during the interaction, one can neglect the contribution of the free atomic Hamiltonian, so that the evolution of the wave function is approximately governed by the following Schr\"odinger equation: 
\begin{equation}
i\hbar\partial_t\Psi_Q=\hat{H}_I\Psi_Q=g(z_a)\hat{l}_{z_e}\Psi_Q. 
\label{Schroedinger equation SG angular momentum interaction}
\end{equation}
For simplicity, let us assume that the initial wave function of the total system prior to entering the Stern-Gerlach magnetic apparatus is separable 
\begin{equation}
\Psi_Q(q_e,q_a;0)=\psi_0(q_e)\varphi_0(q_a). 
\label{initial wave function Stern-Gerlach measurement}
\end{equation}
Let us further expand the total wave function at time $t$, $\Psi_Q(q_e,q_a;t)$, in term of the complete set of orthonormal eigenfunctions of the $z-$angular momentum operator $\{\phi_{l_z}(q_e)\}$, $l_z=1,2,\dots$, satisfying $\hat{l}_{z_e}\phi_{l_z}=\omega_{l_z}\phi_{l_z}$, where $\omega_{l_z}$ is the corresponding eigenvalue. One thus has 
\begin{equation}
\Psi_Q(q_e,q_a;t)=\sum_{l_z}c_{l_z}\phi_{l_z}(q_e)\varphi_{l_z}(q_a;t), 
\label{wave function SG angular momentum}
\end{equation} 
where $\{c_{l_z}\}$ is a set of complex numbers. Putting $t=0$, one must identify $\varphi_{l_z}(q_a;0)=\varphi_0(q_a)$ for all $l_z$, and $\psi_0(q_e)=\sum_{l_z}c_{l_z}\phi_{l_z}(q_e)$. Hence $\{c_{l_z}\}$ is the set of coefficients of expansion of the initial electronic wave function in term of the complete set of orthonormal $z-$angular momentum eigenfunctions. 

Inserting Eq. (\ref{wave function SG angular momentum}) into the Schr\"odinger equation of (\ref{Schroedinger equation SG angular momentum interaction}) one obtains $ 
i\hbar\partial_t\varphi_{l_z}=g(z_a)\omega_{l_z}\varphi_{l_z}$, which can then be directly integrated to give 
\begin{equation}
\varphi_{l_z}(q_a;t)=\varphi_0(q_a)\exp(-i \mu \omega_{l_z}tz_a/\hbar).
\end{equation}
Putting this back into Eq. (\ref{wave function SG angular momentum}), the total wave function at the exit of the Stern-Gerlach magnetic system at $t=T$ is thus  
\begin{eqnarray}
\Psi_Q(q_e,q_a;T)=\sum_{l_z}c_{l_z}\phi_{l_z}(q_e)\varphi_0(q_a)e^{-\frac{i}{\hbar}\Delta_{l_z}z_a},
\label{superposition state SG angular momentum}
\end{eqnarray} 
where we have defined $\Delta_{l_z}\doteq \mu \omega_{l_z} T$. One can thus see that as time flows, the atomic wave function gets a new phase with {\it a wave vector} along the $z-$direction given by $\Delta_{l_z}$ and becomes {\it entangled} with the electronic wave function. 

After passing through the Stern-Gerlach magnet, at $t\ge T$, the wave function, now denoted by $\Psi_M(q;t_M)$, where $t_M=t-T\ge 0$, then evolves approximately freely. Let us take the direction of the beam of the atoms as the $y-$axis, so that the $y-$degree of freedom of the evolution of the wave function is not relevant for our subsequence discussion. After passing through the magnetic field, the wave function is thus governed by the following Schr\"odinger equation:
\begin{equation}
i\hbar\partial_t\Psi_M=\hat{H}_a\Psi_M=-\frac{\hbar^2}{2m_a}\big(\partial_{x_a}^2+\partial_{z_a}^2\big)\Psi_M. 
\label{free atom Schroedinger}
\end{equation}
The above equation has to be solved subject to the initial wave function at $t_M=0$ given by Eq. (\ref{superposition state SG angular momentum}): $\Psi_M(q;0)=\Psi_Q(q;T)$. To do this, let us take a concrete model when the initial atomic wave function is given by a separable Gaussian 
\begin{equation}
\varphi_0(x_a,z_a)\sim\exp\Big(-\frac{x_a^2}{4\sigma_{x_0}^2}\Big)\times\exp\Big(-\frac{z_a^2}{4\sigma_{z_0}^2}\Big),
\label{initial Gaussian}
\end{equation}
up to a normalization constant, where $\sigma_{x_0}$ and $\sigma_{z_0}$ are the widths of the Gaussians for the $x-$ and $z-$degrees of freedom respectively. The Schr\"odinger equation of (\ref{free atom Schroedinger}) can then be solved exactly to give 
\begin{eqnarray}
\Psi_M(q;t_M)\sim\sum_{l_z}c_{l_z}\phi_{l_z}(q_e)\varphi_{l_z}(q_a;t_M),\hspace{0mm}\nonumber\\
\mbox{with}\hspace{2mm}\varphi_{l_z}(q_a;t_M)=e^{-\frac{x_a^2}{4\sigma_{x_t}\sigma_{x_0}}}\hspace{20mm}\nonumber\\
\times e^{-\frac{(z_a-\frac{\Delta_{l_z}}{m_a}t_M)^2}{4\sigma_{z_t}\sigma_{z_0}}-i\frac{\Delta_{l_z}}{\hbar}(z_a-\frac{1}{2}\frac{\Delta_{l_z}}{m_a}t_M)},
\label{final wave function}  
\end{eqnarray} 
where $\sigma_{i_t}=\sigma_{i_0}(1+i\hbar t_M/2m_a\sigma_{i_0}^2)$, $i=x,z$. 

One can thus see that at time $t_M=t-T$, measured just after the particle leaving the region with non-vanishing magnetic field of the Stern-Gerlach apparatus, the $z-$part of the Gaussian packet becomes a series of Gaussian packets, the center of each is shifted from that of the original with an amount that depends linearly on $\omega_{l_z}$ 
\begin{equation}
\frac{\Delta_{l_z}}{m_a}t_M=\frac{\mu \omega_{l_z} T t_M}{m_a},
\label{center of Gaussian z} 
\end{equation}
while the center of the $x-$part of the atomic Gaussian wave function remains fixed; the widths of all Gaussians are increasing with time. It is also clear that the sign of the shift is the same as the sign of $\omega_{l_z}$. Each Gaussian packet $\varphi_{l_z}(q_a;t_M)$ is correlated to one of the eigenfunction $\phi_{l_z}(q_e)$ of the $z-$angular momentum operator $\hat{l}_{z_e}$ corresponding to eigenvalue $\omega_{l_z}$. Further, the distance between the centers of two neighboring packets is given by
\begin{equation}
\delta_{l_z}=\frac{\mu T t_M}{m_a}(\omega_{l_z}-\omega_{l_z-1}), 
\label{center of Gaussians}
\end{equation}
which is larger for larger values of $\mu$, $T$ and $t_M$. If this distance is much larger than the $z-$part spatial spreading of the Gaussian wave packets $\varphi_{l_z}(q_a;t_M)$, then the $z-$part of the series of Gaussians are effectively not overlapping with each other, each is correlated, one to one, to the eigenfunctions of the $z-$angular momentum operator $\phi_{l_z}(q_e)$.  

Next, to have a physically and operationally smooth quantum-classical correspondence, one must let $q_a(t_M)$ has the same {\it physical} and {\it operational} status as the underlying classical mechanical system discussed in the previous subsection: namely, it must be regarded as the pointer of the measurement, the reading of the experiment. One may then {\it infer} that the `outcome' of a single measurement event corresponds to the packet $\varphi_{l_z}(q_a;t_M)$ whose support is actually entered by the atom. Namely, if $z_a(t_M)$ belongs to the spatially localized $z-$part support of $\varphi_{l_z}(q_a;t_M)$, then we {\it operationally} admit (register) that the result of the measurement is given by $\omega_{l_z}$, the eigenvalue of $\hat{l}_{z_e}$ whose corresponding eigenfunction $\phi_{l_z}(q_e)$ is correlated with $\varphi_{l_z}(q_a;t_M)$. The probability that the measurement yields $\omega_{l_z}$ is thus equal to the relative frequency that $z_a(t_M)$ enters the support of $\varphi_{l_z}(q_a;t_M)$ in a large (in principle infinite) number of identical experiments.  

\subsection{Ensemble of identical measurement and Born's rule}
  
It is then imperative to calculate the probability that $z_a(t_M)$ belongs to the support of $\varphi_{l_z}(q_a;t_M)$ when the initial electronic wave function is given by $\phi(q_e)=\sum_{l_z}c_{l_z}\phi_{l_z}(q_e)$. First, given the total wave function at time $t_M$, $\Psi_M(t_M)$, the probability that the measurement yields $\omega_{l_z}$ can be written as, following Eq. (\ref{pre Born's rule hvm SMQ}),  
\begin{eqnarray}
P(\omega_{l_z}|b_z,\Psi_M(b_z))=\int dq_edq_a P(\omega_{l_z}|q_e,q_a,b_z,\Psi_M(b_z))\nonumber\\
\times P(q_e,q_a|b_z,\Psi_M(b_z)), 
\label{Born's rule hvm SMQ} 
\end{eqnarray}  
where $b_z$ is the unit vector along the $z-$axis, $P(q_e,q_a|b_z,\Psi_M(b_z))$ is the joint-probability of $q_e$ and $q_a$ when the total wave function is $\Psi_M(b_z)$ at time $t_M$ and $P(\omega_{l_z}|q_e,q_a,b_z,\Psi_M(b_z))$ is the probability that the measurement yields $\omega_{l_z}$ when the configuration is $q=(q_e,q_a)$ and the total wave function is $\Psi_M(b_z)$ at time $t_M$. Here we have made transparent the dependence of all the probabilities on $b_z$ (the magnitude of the magnetic field plays marginal role in the model). 

Comparing Eq. (\ref{Born's rule hvm SMQ}) with Eq. (\ref{pre Born's rule hvm SMQ}), we have thus regarded the configuration of the system, $q=(q_e,q_a)$, as the hidden variable in the sense discussed in Section \ref{motivation}: $\lambda=(q_e,q_a)$. Such an identification is natural, since, as is evident in Eq. (\ref{fundamental equation rederived}), it is only by specifying the configuration of the system that the wave function defined in Eqs. (\ref{general wave function}) or (\ref{quantum wave function}) determines the momentum or velocity of the system. Moreover, notice that unlike Eq. (\ref{pre Born's rule hvm SMQ}) which is supposed to depend only on the wave function at the preparation $\psi_0$, Eq. (\ref{Born's rule hvm SMQ}) takes into account the fact that as is evident in the discussion in the previous subsection, the whole wave function of the system+apparatus plays irreducible role. We have also made explicit the fact that, due to the fluctuation of infinitesimal stationary action, by construction, the total wave function after the interaction must depend on the setting parameter $b_z$. 

Let us now discuss the two quantities that appear on the right hand of Eq. (\ref{Born's rule hvm SMQ}). First, one has, inserting Eq. (\ref{final wave function}) into Eq. (\ref{Born's statistical interpretation}),  
\begin{eqnarray}
P(q_e,q_a|b_z,\Psi_M(b_z))=\rho_Q(q;t_M,b_z)=|\Psi_M(q;t_M,b_z)|^2\nonumber\\
=\sum_{l_z}|c_{l_z}(b_z)|^2|\phi_{l_z}(q_e;b_z)|^2|\varphi_{l_z}(q_a;t_M,b_z)|^2, 
\label{hv conditioned on Psi z}
\end{eqnarray} 
where in the last (approximate) equality we have taken into account the fact that for sufficiently large values of $\mu$ and $T$, $\{\varphi_{l_z}(q_a;t_M)\}$ in Eq. (\ref{final wave function}) effectively does not overlap for different values of $l_z$ so that the cross-terms are all (approximately) vanishing, and for later purpose, we have made explicit the dependence of all terms on $b_z$. 

Further, since as argued above, given the configuration of the whole system and the total wave function, the outcome of measurement is operationally inferred to be $\omega_{l_z}$ if $z_a(t_M)$ belongs to the support of $\varphi_{l_z}(q_a;t_M,b_z)$, then $P(\omega_{l_z}|q_e,q_a,b_z,\Psi_M(b_z))$ is given by 
\begin{eqnarray}
P(\omega_{l_z}|q_e,q_a,b_z,\Psi_M(b_z))={\bf 1}\{z_a(t_M;b_z)\in\mathcal{U}_{\varphi_{l_z}}\},
\label{outcome conditioned on hv z} 
\end{eqnarray}   
where $\mathcal{U}_{\varphi_{l_z}}$ is the support of $\varphi_{l_z}(q_a;t_M,b_z)$ at time $t_M$, and following the notation of Ref. \cite{Gill notation}, ${\bf 1}\{\mbox{``event''}\}$ is an indicator variable which is equal to $1$ if the ``event'' happens and $0$ if not. 
     
Finally, inserting Eqs. (\ref{hv conditioned on Psi z}) and (\ref{outcome conditioned on hv z}) into Eq. (\ref{Born's rule hvm SMQ}) one gets
\begin{eqnarray}
P(\omega_{l_z}|b_z,\Psi_M(b_z))=\int dq_edq_a|c_{l_z}(b_z)|^2|\phi_{l_z}(q_e;b_z)|^2\nonumber\\
\times |\varphi_{l_z}(q_a;b_z)|^2=|c_{l_z}(b_z)|^2,
\label{Born's rule}
\end{eqnarray}
reproducing the prediction of standard quantum mechanics as prescribed by the Born's rule. 
 
From the isotropy of the classical Hamiltonian and the fact that the model is coordinate free, the same conclusion must apply for any orientation of the magnetic field. For example, instead of directing the magnetic field of the Stern-Gerlach apparatus along the $z-$axis, let us now align the magnetic field along, say the $x-$axis. In this case, one has $B=(B_x,0,0)$ where $B_x=B'x_a$ so that the interaction Hamiltonian now takes the form \begin{equation}
H_I=\frac{e}{2m_e^{\xi}c}B\cdot l_e\approx g_{\xi}(x_a)l_{x_e}, 
\label{Hamiltonian SG x}
\end{equation}
where $l_{x_e}=y_ep_{z_e}-z_ep_{y_e}$ is the $x-$component angular momentum of the electron. Applying the statistical model, neglecting the free Hamiltonian of the electron and repeating exactly all the steps for the case when the magnetic field is pointing along the $z-$axis, one immediately obtains, in the lowest order approximation
\begin{eqnarray} 
i\hbar\partial_t\Psi_Q=(\hat{H}_I+\hat{H}_a)\Psi_Q,\hspace{0mm}\nonumber\\
\hat{H}_I=g(x_a)\hat{l}_{x_e},\hspace{2mm}\hat{H}_a=\frac{\hat{p}_a^2}{2m_a}, 
\label{Schroedinger equation SG angular momentum x-part}
\end{eqnarray}
where ${\hat l}_{x_e}\doteq y_e\hat{p}_{z_e}-z_e\hat{p}_{y_e}$ is the quantum mechanical $x-$angular momentum operator pertaining to the electron.

One can then proceed exactly as for the case when the Stern-Gerlach magnet is pointing along the $z-$axis to describe a single measurement event and the statistics of outcomes of its ensemble. First, evolving the initial separable wave function of Eq. (\ref{initial wave function Stern-Gerlach measurement}), neglecting the free atomic Hamiltonian during the interaction, the atomic part of the wave function gets a new phase {\it now} with a wave vector along the $x-$axis, and it becomes {\it entangled} with the electronic wave function
\begin{eqnarray}
\Psi_Q(q_e,q_a;T,b_x)=\sum_{l_x}c_{l_x}(b_x)\phi_{l_x}(q_e;b_x)\varphi_0(q_a)e^{-\frac{i}{\hbar}\Delta_{l_x}x_a},
\label{superposition state SG angular momentum x} 
\end{eqnarray} 
where $\phi_{l_x}$, $l_x=0,1,2,\dots$, is the eigenfunction belonging to the eigenvalue $\omega_{l_x}$ of the $x-$angular momentum operator $\hat{l}_x$, $\hat{l}_x\phi_{l_x}=\omega_{l_x}\phi_{l_e}$, $\{c_{l_x}(b_x)\}$ is the set of coefficients of expansion of the initial electronic wave function in terms of the complete orthonormal set the $x-$angular momentum eigenfunctions, $\{\phi_{l_x}\}$, and $\Delta_{l_x}\doteq \mu \omega_{l_x} T$. 

Further, after passing through the magnetic region of the Stern-Gerlach apparatus, the total wave function evolves according to the free atomic Hamiltonian. Assuming again that the initial atomic wave function is given by the same Gaussian of Eq. (\ref{initial Gaussian}), one obtains at $t_M=t-T$ 
\begin{eqnarray}
\Psi_M(q;t_M,b_x)\sim\sum_{l_x}c_{l_x}(b_x)\phi_{l_x}(q_e;b_x)\varphi_{l_x}(q_a;t_M,b_x),\hspace{0mm}\nonumber\\
\mbox{with}\hspace{2mm}\varphi_{l_x}(q_a;t_M,b_x)=e^{-\frac{z_a^2}{4\sigma_{z_t}\sigma_{z_0}}}\hspace{20mm}\nonumber\\
\times e^{-\frac{(x_a-\frac{\Delta_{l_x}}{m_a}t_M)^2}{4\sigma_{x_t}\sigma_{x_0}}-i\frac{\Delta_{l_x}}{\hbar}(x_a-\frac{1}{2}\frac{\Delta_{l_x}}{m_a}t_M)}. 
\label{final wave function x}  
\end{eqnarray} 
The $x-$part of the Gaussian packet becomes a series of Gaussian packets the center of each is shifted from that of the original with an amount that is linear with $\omega_{l_x}$, while the center of the $z-$part Gaussian is left unchanged. Repeating all the calculation as before, assuming that the $x-$part of the series of Gaussians atomic wave function $\varphi_{l_x}$ are (approximately) not overlapping with each other for sufficiently large $\mu$ and $T$, the distribution of the hidden variables $(q_e,q_a)$ at time $t_M$ now takes the form, due to Eq. (\ref{Born's statistical interpretation}), 
\begin{eqnarray} 
P(q_e,q_a|b_x,\Psi_M(b_x))=\rho_Q(q;t_M,b_x)=|\Psi_M(q;t_M,b_x)|^2\hspace{0mm}\nonumber\\
=\sum_{l_x}|c_{l_x}(b_x)|^2|\phi_{l_x}(q_e;b_x)|^2|\varphi_{l_x}(q_a;t_M,b_x)|^2. 
\label{hv conditioned on Psi SG x}
\end{eqnarray}  

On the other hand, to have a smooth quantum-classical correspondence, regarding $q_a(t)$ as the pointer to infer the measurement outcome as before, one has 
\begin{equation}
P(\omega_{l_x}|q_e,q_a,b_x,\Psi_M(b_x))={\bf 1}\{x_a(t_M;b_x)\in\mathcal{U}_{\varphi_{l_x}}\}, 
\label{conditional probability of outcome x}
\end{equation}    
namely, the experiment registers $\omega_{l_x}$ if $x_a(t_M;b_x)$ belongs to the support of $\varphi_{l_x}(q_a;t_M,b_x)$ denoted by $\mathcal{U}_{\varphi_{l_x}}$. 

Finally, from Eqs. (\ref{hv conditioned on Psi SG x}) and (\ref{conditional probability of outcome x}), one obtains
\begin{eqnarray}
P(\omega_{l_x}|b_x,\Psi_M(b_x))=\int dq_edq_aP(\omega_{l_x}|q_e,q_a,b_x,\Psi_M(b_x))\nonumber\\
\times P(q_e,q_a|b_x,\Psi_M(b_x))=|c_{l_x}(b_x)|^2,  
\label{Born's rule x}
\end{eqnarray}
reproducing again the prediction of quantum mechanics as prescribed by the Born's rule.

Hence, in general, for any orientation of the magnetic field denoted by the unit vector $b$, one regains the prediction of quantum mechanics. That is, each single measurement yields $\omega_{l_b}$, $l_b=1,2,\dots$, one of the eigenvalues of the quantum mechanical angular momentum operator $\hat{l}_b=(q\times\hat{p})\cdot b$, and repeating the experiment with identical preparation represented by the initial electronic wave function $\psi_0(q_e)$, the relative frequency to get outcome $\omega_{l_b}$ is given by the Born's rule
\begin{eqnarray}
P(\omega_{l_b}|b,\Psi_M(b))=|c_{l_b}(b)|^2,  
\label{Born's rule b}
\end{eqnarray} 
where $c_{l_b}=\int dq_e\phi_{l_b}^*(q_e)\psi_0(q_e)$ with $\phi_{l_b}$ is the eigenfunction of $\hat{l}_b$ belonging to the eigenvalue $\omega_{l_b}$.  

\subsection{Discussion}

First, let us note that the above argumentation is similar to that of the de Broglie-Bohm hidden variable model, the pilot-wave theory \cite{Bohmian mechanics}, in which the configuration of the whole system is also regarded as the hidden variable. One can see from Eqs. (\ref{outcome conditioned on hv z}) and (\ref{conditional probability of outcome x}) that like the pilot-wave theory, the model is deterministic in the sense that the hidden variables determine the outcome of measurement: $\omega=\omega(q_a;b)$. However, unlike the pilot-wave theory, the wave function defined in Eqs. (\ref{general wave function}) or (\ref{quantum wave function}) is evidently {\it not} a physical field, but is an artificial mathematical construct. The fundamental assumption that the wave function in pilot-wave theory is a {\it physical field} living in {\it configuration space} rather than in ordinary space has been known to lead to a conceptual difficulty, and furthermore implies rigid nonlocality. By contrast, the present model satisfies the separability condition of Eq. (\ref{principle of Local Causality}) so that it is objectively local: what happens in some region of spacetime cannot have any influence whatsoever outside its future light cone. In this sense, the upper equation in (\ref{fundamental equation rederived}) can {\it not} be regarded as a causal-dynamical guidance relation as in pilot-wave theory, rather it is a kinematical relation. Moreover, unlike pilot-wave theory in which the quantum dynamics and kinematics are postulated, and so is the additional guidance relation, in the statistical model, they are derived from first principle.   
 
As argued at the end of subsection \ref{locality and inseparability} and subsection \ref{entanglement}, since action is evaluated in configuration space instead of in ordinary space, then the random fluctuation of infinitesimal stationary action with respect to the whole `system+apparatus' is {\it not} separable into the fluctuation pertaining to the system and that pertaining to the apparatus. This is explicitly reflected in the {\it inseparability} of wave function of Eqs. (\ref{superposition state SG angular momentum}) or (\ref{superposition state SG angular momentum x}). Like the corresponding classical model, here the atomic degree of freedom plays the role as the apparatus. The total wave function becomes {\it entangled} due to interaction. The whole system+apparatus must then be regarded as a single unanalyzable whole, both fluctuates together inseparably. Hence, unlike measurement in classical mechanics discussed in the previous section in which the interaction Hamiltonian conserves the relevant component of the angular momentum of the particle being measured, in the statistical model, the same component of the angular momentum prior to measurement is inevitably disturbed. 

The role of the apparatus is no longer merely revealing the value of the angular momentum prior to measurement-interaction. In this sense, the term ``apparatus'' thus loses its significant (classical) role. It is even questionable to call such an experiment a measurement. There are however at least two reasons to call such an experiment a measurement: first, its classical limit can unambiguously be regarded as measurement, and second, as argued in Ref. \cite{AgungSMQ7}, repeating immediately the same measurement as the previous one yields the same outcome. It has been shown in Ref. \cite{AgungSMQ7}, however, that using Eq. (\ref{Born's rule b}) and the upper equation in (\ref{fundamental equation rederived}), the average of the {\it outcomes} of measurement when the magnetic field is directed along an axis in an ensemble of identical experiments, is equal to the average of the {\it actual values} of the angular momentum along the same axis {\it prior} to measurement over the distribution of the configuration. 
 
The orientation of the magnetic field of the Stern-Gerlach apparatus evidently plays essential role in the measurement of angular momentum. First, it determines the form of the interaction Hamiltonian which in turn is responsible for the emergence of a mathematical entity $\hat{l}_{b}$, which in the standard quantum mechanical nomenclature is called as the angular momentum operator along the $b-$axis. As shown above, the latter is central in the subsequence discussion, determining the set of all possible outcomes of each single measurement event that is given by its discrete spectrum of eigenvalues, and the probability to get various outcomes in an ensemble of identical measurements that is given by the absolute square of the overlap between the initial electronic wave function and the eigenfunction of $\hat{l}_{b}$, that is the Born's rules.   

It is then clear from the above discussion that the quantum mechanical angular momentum operators $\hat{l}_b$ arise formally as {\it artificial} mathematical entities depending on the {\it whole} configuration/arrangement of the measurement-interaction described by the interaction Hamiltonian. Hence, they can{\it not} be granted independent physical ontology attached to the system being measured alone. Since the outcome of measurement is given by one of the eigenvalues of the operator, then one may conclude that the outcome of the measurement can{\it not} be regarded as reflecting the property attached to the system being measured {\it alone}, but brings with it the whole configuration of the measurement-interaction determined by the apparatus setting. It is also clear that $\hat{l}_b$ for different $b$ arise in mutually exclusive (incompatible) contexts (arrangements) so that one cannot manipulate the eigenvalues of, say a pair of different $\hat{l}_b$, freely. The statistical model thus provides an explicit manifestation of the Bohr's complementarity principle \cite{Bohr complementarity,Howard on Bohr}, yet interestingly within Einstein's local realist world view. The origin of the above feature can of course again be traced back to the fact that in the statistical model, the origin of the randomness is due to the random fluctuation of the infinitesimal stationary action, which, for an interacting compound (system+apparatus), can{\it not} be separated into the the fluctuation with respect to each subsystem.   
 
We have thus given a causal description for the emergence of Born's rule within the local-causal statistical model in which the configuration of the whole system+apparatus is regarded as the hidden variables in the sense of Section \ref{motivation}, which together with the whole wave function determine the measurement outcome.  
  
\section{Bell's locality assumption}

We have argued that measurement of angular momentum can be described as a specific type of physical interaction. Namely, the cases when there is no measurement and when there is a measurement are treated in a unified way within the statistical model, satisfying the same local-causal dynamical and statistical law given by the exponential distribution of deviation from infinitesimal stationary action of Eq. (\ref{exponential distribution of DISA}), necessitating no external concepts. 

Now let us consider the situation in Bell-type experiment where one is interested in the statistical correlation between a set of pairs of spacelike separated detection (measurement) events.  Since the statistics of {\it any} events, whether they are detection events in measurement-interaction or not, must satisfy the {\it objective} locality (separability) condition of Eq. (\ref{principle of Local Causality}) when they are separated by spacelike interval, then the statistics of the measurement outcomes must necessarily satisfy the Bell's locality assumption of Eq. (\ref{SH}). That is to say, if at the two wings of the Bell-type experiment we perform Stern-Gerlach experiments, then the probability that at the end of a measurement one of the atoms in the pair at the first(second) wing (labeled below by subscript 1(2)) with coordinate $q_{a_{1(2)}}$ enters the support of a particular packet $\varphi_{l_{b_{1(2)}}}$ corresponding to a certain measurement outcome $\omega_{l_{b_{1(2)}}}$ with $l_{b_{1(2)}}=1,2,3,\dots$, when the orientation of the corresponding Stern-Gerlach magnet is $b_{1(2)}$, is independent from anything that happens with the other atom at the second(first) wing, and vice versa 
\begin{eqnarray}
P(\omega_{l_{b_1}},\omega_{l_{b_2}}|\{q_{e_1},q_{a_1},b_1\},\{q_{e_2},q_{a_2},b_2\},\psi_0)\hspace{20mm}\nonumber\\
={\bf 1}\{q_{a_1}(t_M;b_1)\in\mathcal{U}_{\varphi_{l_{b_1}}}\}{\bf 1}\{q_{a_2}(t_M;b_2)\in\mathcal{U}_{\varphi_{l_{b_2}}}\}\nonumber\\
=P(\omega_{l_{b_1}}|q_{e_1},q_{a_1},b_1,\psi_0)P(\omega_{l_{b_2}}|q_{e_2},q_{a_2},b_2,\psi_0). \hspace{0mm} 
\end{eqnarray} 
This is so even when the pair of the atoms are emitted from a source in any entangled state $\psi_0$. Otherwise, if the above equation is not valid, since the effective velocity is given by $\widetilde{\dot{q}_{a_i}}=\partial_{q_{a_i}}S_Q/m_i$ (see Eq. (\ref{effective classical velocity field SG angular momentum})), then the phase $S_Q$ must be indecomposable so that its differential is also indecomposable contradicting the condition of statistical separability of Eq. (\ref{principle of Local Causality}). 
    
The violations of Bell inequalities using orbital angular momentum of entangled photons have  been demonstrated experimentally, confirming accurately the prediction of  quantum mechanics \cite{experiment oam ES}. Assuming that the local-causal statistical model applies as well to the dynamics and statistics of outcomes of the measurement of orbital angular momentum of photon reproducing the predictions of quantum mechanics, noting that the model supports entangled states and explicitly satisfies the Bell's locality assumption, then the quantum mechanical and experimental violation of Bell's inequalities must have noting to do with nonlocality.  

\section{Conclusion}

It is {\it widely} believed that Bell's no-go theorem has put a strong argument that some of the predictions of quantum mechanics which are obtained by applying the {\it Born's rule} to a class of {\it entangled states} are indescribable within {\it any} statistical model supposed to complete the description of quantum mechanics by adding the wave function with a set of hidden variables, if the model is locally causal \cite{Bell paper,CHSH inequality,CH stochastic model,Bell's theorem general ES}. In fact, in formulating his theorem, Bell was much inspired by the pilot-wave theory which gives an example of a completion of the description of quantum mechanics by regarding the configuration of the system as the hidden variable, and pilot-wave theory is rigidly nonlocal. To show this, Bell and others have devised certain statistical inequalities, the Bell's inequalities, and argued that they must be satisfied by any local statistical model satisfying the Bell's locality assumption of Eq. (\ref{SH}). The inequalities are violated by the statistical predictions of quantum mechanics, and numerous experiments have been carried out showing results in favor of quantum mechanics \cite{Brunner review}.      

On the other hand, we have shown that quantum dynamics and kinematics, including the entangled states, can be derived from a statistical model based on a random fluctuation of infinitesimal stationary action which is singled out uniquely by the principle of Locality. The origin of the generation of entanglement is argued as due to the fact that for a compound with interacting subsystems, the fluctuation of infinitesimal stationary action of the whole compound cannot be fundamentally separated into the fluctuation of each subsystems. We have also shown, by applying the model to the Stern-Gerlach experiment for the measurement of angular momentum, that Born's rule emerges as a mathematical theorem of the local-causal model. Measurement is thus treated in equal footing as the other types of interaction necessitating no external concepts. This led us to argue that the objective locality of the model of Eq. (\ref{principle of Local Causality}) implies the Bell's locality assumption of Eq. (\ref{SH}) for spacelike separated joint-measurement events. 

Hence, since entanglement and Born's rule are responsible for the quantum mechanical violation of Bell's inequalities and the former are emergent within the local-causal statistical model satisfying Bell's locality assumption, then nonlocality must {\it not} be blamed as the source of violation. Moreover, since the predictions of quantum mechanics is confirmed very accurately by numerous experiments, neglecting all the complexities that might arise due various potential experimental loopholes, one may conclude that Nature does {\it not} use nonlocality to violate the Bell's inequalities.


\begin{thebibliography}{10}      
 
\bibitem{Bell paper} J. S. Bell, Physics 1, 195 (1964); Speakable and Unspeakable in Quantum Mechanics, Cambridge University Press, Cambridge, 1987.  
\bibitem{CHSH inequality} J. F. Clauser, M. A. Horne, A. Shimony, R. A. Holt, Phys. Rev. Lett. 23, 880 (1969).
\bibitem{CH stochastic model} J. Clauser, M. Horne, Phys. Rev. D 10, 526 (1974).
\bibitem{Bell's theorem general ES} N. Gisin and A. Peres, Phys. Lett. A 162, 15 (1992). 

\bibitem{Jarret PI and OI} J. Jarrett, Nous 18, 569 (1984).
\bibitem{Shimony PI and OI} A. Shimony in J. Ellis and D. Amati (eds.), Quantum Reflections, Cambridge University Press, Cambridge, 2000.  

\bibitem{Bell free will} J. S. Bell, Epistemol. Lett. 9, 11 (1976).
\bibitem{Shimony free will} A. Shimony, M. A. Horne, J. S. Clauser, Epistemol. Lett. 13, 9 (1976).  
\bibitem{Espagnat free will}B. d' Espagnat, Phys. Rep. 110,  201 (1984).
\bibitem{Shimony model} A. Shimony, M. A. Horne and J. F. Clauser, Dialectica 39, 97 (1985). 
\bibitem{Brans full causal} C. Brans, Int. J. Theor. Phys. 27, 219 (1988). 
\bibitem{Kofler free will relaxation} J. Kofler, T. Paterek and C. Brukner, Phys. Rev. A 73, 022104 (2006).
\bibitem{Hall free will relaxation} M. J. W. Hall, Phys. Rev. Lett. 105, 250404 (2010); Phys. Rev. A 84, 022102 (2011).   
\bibitem{Di Lorenzo MI} A. Di Lorenzo, Journal of Physics A: Mathematical and Theoretical 45, 265302 (2012).
\bibitem{Vervoort} L. Vervoort, Found. Phys. 43, 769 (2013).

\bibitem{Vorobev theorem} N. N. Vorob'ev, Theory of Probability and its Applications VII, 147-162 (1962). 

\bibitem{Accardi contextual loophole} L. Accardi, Phys. Rep. 77, 169 (1981). 
\bibitem{Pitowsky contextual loophole} I. Pitowsky, ``From George Boole to John Bell: The Origins of Bell's Inequalities,'' in M. Kafatos (Ed.), Proc. Conf. Bell's Theorem, Quantum theory and Conceptions of the Universe, Kluwer, Dordrecht, pp. 37-49, 1989; Brit. J. Phil. Sci. 45, 95 (1994).  
\bibitem{de la Pena contextual loophole} A. M. Cetto, T. Brody and L. de la Pen\~a, Lett. Nuovo Cimento 5, 177 (1997). 
\bibitem{Khrennikov contextual loophole} A. Y. Khrennikov, Found. Phys. 32, 1159 (2002); Contexual Approach to Quantum Formalism, Springer, Berlin (2009); arXiv:0709.3909v2.  
\bibitem{Volovich contextual loophole} I. V. Volovich, in A. Y. Khrennikov (ed.), Proc. Conf. Quantum Theory: Reconsideration of Foundations. Ser. Math. Modeling, vol 2, p. 423. V\"axj\"o University Press, V\"axj\"o (2002). 
\bibitem{Hess-Philips} K. Hess and W. Philipp: ``Bell's Theorem: Critique of Proofs With And
Without Inequalities,'' in A. Y. Khrennikov (ed.), Proc. Conf. Foundations of Probability and Physics-3. AIP Conference Proceedings, vol. 750, pp. 150-155. AIP, New York (2005); arXiv:quant-ph/0410015v1.  
\bibitem{Hess contextual loophole} K. Hess, K. Michielsen and H. De Raedt, Europhys. Lett. 87, 60007 (2009).   
\bibitem{Nieuwenhuizen loopholes} T. M. Niuewenhuizen, Found. Phys. 41, 580 (2011).
    
\bibitem{AgungSMQ6} A. Budiyono, Int. J. Theor. Phys. 52, 4237 (2013).
\bibitem{AgungSMQ8} A. Budiyono, Physica A 399, 40 (2014).  
\bibitem{AgungSMQ4} A. Budiyono, Physica A 392, 307 (2013).
\bibitem{AgungSMQ7} A. Budiyono, J. Stat. Mech.: Theory and Experiment P11007, 1 (2013). 

\bibitem{Brunner review} For a recent review, see N. Brunner, D. Cavalcanti, S. Pironio, V. Scarani, and S. Wehner, arXiv:1303.2849v1.  

\bibitem{Ellis top-down} G. F. R. Ellis, Interface Focus 2, 126 (2012). 

\bibitem{some complication} Such a setting does not exactly satisfy Maxwell equation. See A. B\"ohm, Quantum Mechanics, Berlin, Springer-Verlag, 1986.  

\bibitem{Gill notation} R. D. Gill, G. Weihs, A. Zeilinger, and M. Zukowski, Proc. Nat. Acad. Sci. 9, 14632 (2002). 

\bibitem{Bohmian mechanics} D. Bohm and B. Hiley, The Undivided Universe: an Ontological Interpretation of Quantum Theory, Routledge, London, 1993.   

\bibitem{Bohr complementarity} N. Bohr, in Quantum Theory and Measurement, in J.A.
Wheeler, W.H. Zurek, (eds.), Princeton Univ. Press, Princeton, pages 949, 1984; Atomic Physics and Human Knowledge, 2010, Dover, New York.
\bibitem{Howard on Bohr} D. Howard, Philosophy of Science 71, 669 (2004).  

\bibitem{experiment oam ES} A. C. Dada, J. Leach, G. S. Buller, M. J. Padgett, and E. Andersson, Nature Physics 7, 677 (2011).  

\end{thebibliography}
\end{document}